\documentclass[11pt]{article}
\usepackage{amsfonts, exscale}
\usepackage{amssymb, amsthm,amsmath}

\date{}
\title{Sobolev inequality for localization of pseudo-relativistic energy}
\author{A.~A.~Balinsky and A.~E.~Tyukov}

\newtheorem{Theorem}{Theorem}
\newtheorem{Lemma}{Lemma}

\newtheorem{Remark}{Remark}
\newtheorem*{Keywords}{Keywords}
\newtheorem*{Abstract}{Abstract}

\newcommand{\reff}[1]{{\rm(\ref{#1})}}

\renewcommand{\leq}{\leqslant}
\renewcommand{\geq}{\geqslant}

\begin{document}

\maketitle{}
\begin{center}
Cardiff School of Mathematics, \\
Cardiff University, \\
Senghennydd Road, \\ Cardiff CF24 4AG \\ United Kingdom
\end{center}

\begin{Abstract}\rm
In this article we present Sobolev-type inequalities for the
localization of pseudo-relativistic energy.
\end{Abstract}

\begin{Keywords}\rm
Sobolev-type inequalities, localization of pseudo-relativistic
energy, scale-space theory, stability of matter
\end{Keywords}

In this article we present Sobolev-type inequalities for the
localization of pseudo-relativistic energy. This is a continuation
of our previous article (\cite{BT}), where a Kato-type inequality
for the same localization has been established. In a seminal paper
by Lieb and Yau (\cite{LY})  some inequalities for
pseudo-relativistic energy localization have been used to
establish the stability of matter. We hope that our results can
also be used  in this area. Another reason for studying such
quadratic forms comes from scale-space theory in image processing.
It has been shown (\cite{FS}) that pseudo-relativistic energy
generates better scale-space theory for ${\mathbb R}^2$. Since
real images are defined on bounded domains we need similar
scale-space theory for such domains. There are two possibilities
to do this. One is to take the square root of the Laplacian
(\cite{DFFP}). The other  possibility is to consider the quadratic
form for an image $f(x)$ defined by
$$
\int\limits_{\Omega \times \Omega} {|f(x)-f(y)|^2 \over |x-y|^{3}}
\, dxdy \, .
$$
The second possibility is  preferable since it is easy to calculate
and explicitly involves image contrast. Applications of results
obtained in this paper for image processing will be studied in a
future article.
\par
Our main result is
\begin{Theorem}\label{t1}
Let $\Omega$ be a bounded domain in ${\mathbb R}^n$ such that
$\Omega = \phi (B_1 )$, where $B_1 \subset {\mathbb R}^n$, $n \geq
2$ is a unit ball and $\phi : B_1 \rightarrow \Omega$ is a
diffeomorphism with
\begin{equation}\label{29}
c_1^{-1} |x-y| \leq |\phi (x) -\phi (y)| \leq c_1 |x-y|
\end{equation}
for some constant $c_1=c_1 (\Omega)>1$. Then there exists a
constant \mbox{$c_2 =c_2 (\Omega)>0$} such that
$$
\int\limits_{\Omega \times \Omega} {|f(x)-f(y)|^2 \over
|x-y|^{n+1}} \, dxdy \geq c_2 \left( \int\limits_{\Omega}
|f(x)|^{2n \over n-1} \, dx \right)^{n-1 \over n} \,
$$
for any continuous function $f: \Omega \rightarrow {\mathbb C}$
with $supp \, f \subset \Omega$.
\end{Theorem}
The proof of Theorem~\ref{t1} will be preceded by the proofs of
two auxiliary lemmas.
\begin{Lemma}\label{l1}
Let  $\Omega \subset {\mathbb R}^{n}$   either satisfies the
conditions of Theorem~\ref{t1} or $\Omega ={\mathbb R}^{n}$. Let
$0 < \gamma <1$. Then there exists a constant $c_3=c_3 (\Omega,
\gamma)>0$ such that for any $V \subset A \subset \Omega $ one has
\begin{equation}\label{15}
\int\limits_{\Omega \setminus A} \int\limits_{V} {dx dy \over
|x-y|^{n+1}} \geq c_3 |V|^{n-1 \over n} \,
\end{equation}
provided that $|A| < \infty$,
\begin{equation}\label{5}
|\Omega \setminus A| >  |A| \,
\end{equation}
and
\begin{equation}\label{17}
|V| \geq \gamma |A| \, .
\end{equation}
\end{Lemma}
\begin{Remark}
The following example shows  the importance of the restriction
that $\Omega=\phi(B_1)$, where the diffeomorphism $\phi$
satisfies~\reff{29}. The domain
$$
\Omega =\{(x_1, x_2) \in {\mathbb R}^2: 0<x_1<1, \, 0< x_2
<x_1^2\}
$$
does not satisfy the conditions of Theorem~\ref{t1}. Let us take
\begin{align*}
A_k &=\{(x_1, x_2) \in {\mathbb R}^2: 0<x_1<{2 \over k}\, , \, 0<
x_2 <x_1^2
\} \, , \\
V_k &=\{(y_1, y_2) \in {\mathbb R}^2: 0<y_1<{1 \over k} \, , \, 0<
y_2 <y_1^2 \} \, .
\end{align*}
Note that   $|\Omega \setminus A_k| >  |A_k|$, $|V_k|= |A_k|/8$
for $k \geq 4$ and so~\reff{5} and~\reff{17} hold. One readily
sees that
$$
|x-y| \geq |x_1-y_1| \geq \left|x_1-{1 \over k}\right| \geq {|x_1|
\over 2 }
$$
for $x=(x_1, x_2) \in \Omega \setminus A_k$, $y=(y_1, y_2) \in
V_k$, where we have used \\ $0<y_1 < 1/k <2/k <x_1$. Hence
$$
\int\limits_{\Omega \setminus A_k} \int\limits_{V_k} {dx dy \over
|x-y|^{3}} \geq  8 |V_k| \int\limits_{\Omega \setminus A_k}
 {dx \over |x_1|^{3}} = {8 \over 3} \, |V_k| \left(1-{2 \over k}
 \right)\, .
$$
Therefore
$$
|V_k|^{-n+1 \over n} \int\limits_{\Omega \setminus A_k}
\int\limits_{V_k} {dx dy \over |x-y|^{3}} \rightarrow 0
$$
as $k \rightarrow \infty$.
\end{Remark}
\begin{proof}[Proof of Lemma~\ref{l1}.]
Let us briefly outline the main ideas of the proof. The proof for
the case $\Omega = {\mathbb R}^n$ is based on the fact that the
expression
$$
|V|^{-n+1 \over n} \int\limits_{{\mathbb R}^n \setminus A}
\int\limits_{V} {dx dy \over |x-y|^{n+1}}
$$
is invariant under the replacement $(A, V) \rightarrow (tA, tV)$,
where $t>0$ is a parameter. In the case $\Omega = B_1$ we
construct sets $\widehat V \subset \widehat A$,  such that $ A
\subset \widehat A$, $ V \subset \widehat V$ and $|\widehat V|
\geq const. |\widehat A | >0$. Then we reduce this case to the
previous one by applying Lemma~\ref{l1} with $(\Omega, A,
V):=({\mathbb R}^n, \widehat A, \widehat V)$. In the most general
case $\Omega=\phi (B_1)$ it is enough to notice that both sides
of~\reff{15} are compatible with the corresponding expressions for
$\Omega = B_1$. Now we proceed to the proof.
\par {\it Case 1.} Let  $\Omega ={\mathbb
R}^n$.
\par Take a set of unit cubes $Q_m \subset {\mathbb R}^n$,  $m \in
{\mathbb N}$ such that ${\mathbb R}^n=\bigcup\limits_{m=1}^\infty
Q_m$. Using that $|x-y| \leq \sqrt{n}$ for $x, y \in Q_m$ we
obtain
\begin{align*}
\int\limits_{A^c}  \int\limits_{V} {dx dy \over |x-y|^{n+1}} &\geq
\sum\limits_{m=1}^\infty \, \, \int\limits_{Q_m \cap A^c} \, \,
\int\limits_{Q_m \cap V} {dx dy \over |x-y|^{n+1}} \\ &\geq {1
\over (\sqrt{n})^{n+1}} \sum\limits_{m=1}^\infty \, \, |Q_m \cap
A^c| |Q_m \cap V| \, ,
\end{align*}
where $A^c={\mathbb R}^n \setminus A$. If  $|A| = 1/2$, then $|Q_m
\cap A^c|\geq 1/2$ and using~\reff{17} we get
\begin{equation}\label{18}
\int\limits_{A^c} \int\limits_{V} {dx dy \over |x-y|^{n+1}} \geq
{1 \over 2 (\sqrt{n})^{n+1} } \, \sum\limits_{m=1}^\infty \, \,
 |Q_m \cap V|={|V| \over 2 (\sqrt{n})^{n+1}}\geq  { \gamma \over 4 (\sqrt{n})^{n+1}}\, .
\end{equation}
In the general case we take $t=(2|A|)^{1/n}$ and $A_0=t^{-1} A$,
$V_0=t^{-1} V$. Clearly $|A_0|=1/2$ and $|V_0| \geq \gamma |A_0|$.
Making the change of variables $x:=tx$, $y:=ty$ and
applying~\reff{18} with $A_0$ and $V_0$ gives
\begin{align*}
\int\limits_{A^c} \int\limits_{V} {dx dy \over |x-y|^{n+1}} =
t^{n-1} \, \int\limits_{A^c_0} \int\limits_{V_0} {dx dy \over
|x-y|^{n+1}} &\geq  t^{n-1} { \gamma \over 4 (\sqrt{n})^{n+1}}
\\
&\geq { \gamma \over 4 (\sqrt{n})^{n+1}} \, |V|^{n-1 \over n} \, ,
\end{align*}
where we used $t^{n-1} \geq (2|A|)^{n-1 \over n} \geq (2|V|)^{n-1
\over n} \geq |V|^{n-1 \over n}$.
\par
{\it Case 2.} We proceed with the case $\Omega=B_1$. We
denote by $B_r$  the ball in ${\mathbb R}^n$ with radius $r$ and
center at the origin. Let $(A_k, V_k)$ be a sequence of sets
minimizing $G=G(A, V)$
$$
G:= {1 \over |V|^{n-1 \over n}} \,  \int\limits_{\Omega \setminus
A} \int\limits_{V} {dx dy \over |x-y|^{n+1}} \, .
$$
among all $(A,V)$ satisfying the conditions of Lemma~\ref{l1}. We
are done if we show that
\begin{equation}\label{28}
G(A_k, V_k) \geq const. > 0
\end{equation}
for all $k \in {\mathbb N}$.
\par
Since $|x-y| \leq 2$ for $x, y \in B_1$,  it follows that
\begin{equation}\label{33}
G \geq 2^{-n-1}|\Omega \setminus A|\,  |V|^{1 \over n} \geq
2^{-n-1}|A|\,  |V|^{1 \over n} \, .
\end{equation}
If   $\inf\limits_{k} |A_k| >0$,  then~\reff{28} is a
straightforward consequence of~\reff{17} and~\reff{33}. Therefore
it suffices to consider the case $\lim\limits_{k \rightarrow
\infty} |A_k|=0$. Without loss of generality we may assume that
for any $k$
\begin{equation}\label{24}
|A_k| <{1 \over 2} \, |B_{1 \over 3} \setminus B_{1 \over 4}| \, .
\end{equation}
\par {\it Case 2a.} Suppose that for all $k$
\begin{equation}\label{26}
|V_k \cap B_{1 \over 2} | > {1 \over 2} \,  |V_k| \, .
\end{equation}
Since $|x-y| \leq 3/2 <3 $ for all $y \in B_{1 \over 2}$,  $x \in
B_1 \setminus B_{3 \over 4}$ and because of~\reff{24}, it follows
that
$$
\int\limits_{(B_1 \setminus B_{3 \over 4}) \setminus A_k} {dx
\over |x-y|^{n+1}} \geq {1 \over 3^{n+1}} \,  |(B_1 \setminus B_{3
\over 4}) \setminus A_k| \geq {1 \over  3^{n+2}} \, |B_1 \setminus
B_{3 \over 4} | =:c_4 >0 \, .
$$
Moreover, for any $y \in B_{1 \over 2}$
$$
\int\limits_{{\mathbb R}^n \setminus B_1} \, {dx \over
|x-y|^{n+1}} \leq \int\limits_{{\mathbb R}^n \setminus B_1} \, {dx
\over (|x|-{1\over 2})^{n+1}}=:c_5 \, .
$$
Thus for all $y \in B_{1 \over 2}$ and $A_k \subset B_1$
satisfying~\reff{24} one has
\begin{equation}\label{25}
\int\limits_{(B_1 \setminus B_{3 \over 4}) \setminus A_k} {dx
\over |x-y|^{n+1}} \geq c_6 \int\limits_{({\mathbb R}^n \setminus
B_{3 \over 4})\setminus A_k} {dx \over |x-y|^{n+1}} \, ,
\end{equation}
where $c_6:=c_4 /(c_5+c_4)$. Using
$$
\int\limits_{B_1 \setminus A_k} =\int\limits_{B_{3 \over 4}
\setminus A_k}+ \int\limits_{(B_1 \setminus B_{3 \over 4})
\setminus A_k}
$$
and~\reff{25} we obtain
\begin{eqnarray*}
\lefteqn{ \int\limits_{B_1 \setminus A_k} \int\limits_{V_k \cap
B_{1 \over 2}} {dx dy \over |x-y|^{n+1}} } \qquad \qquad \qquad
\\
&\geq& \int\limits_{B_{3 \over 4} \setminus A_k} \int\limits_{V_k
\cap B_{1 \over 2}} {dx dy \over |x-y|^{n+1}}+ c_6
\int\limits_{({\mathbb R}^n \setminus B_{3 \over 4}) \setminus
A_k} \int\limits_{V_k \cap B_{1 \over 2}} {dx dy \over
|x-y|^{n+1}}
\\
&\geq& c_6 \int\limits_{{\mathbb R}^n \setminus A_k}
\int\limits_{V_k \cap B_{1 \over 2}} {dx dy \over |x-y|^{n+1}} \,
.
\end{eqnarray*}
Consequently we have
\begin{align*}
\int\limits_{B_1 \setminus A_k} \int\limits_{V_k} {dx dy \over |x
-y|^{n+1}} \geq \int\limits_{B_1 \setminus A_k} \int\limits_{V_k
\cap B_{1 \over 2}} {dx dy \over |x-y|^{n+1}} & \geq c_6
\int\limits_{{\mathbb R}^n \setminus A_k} \int\limits_{V_k \cap
B_{1 \over 2}} {dx dy \over |x-y|^{n+1}} \, .
\end{align*}
Using~\reff{26} we apply Lemma~\ref{l1}  with $(\Omega, A,
V):=({\mathbb R}^n, A_k,V_k \cap B_{1 \over 2})$ to get
$$
\int\limits_{{\mathbb R}^n \setminus A_k} \int\limits_{V_k \cap
B_{1 \over 2}} {dx dy \over |x-y|^{n+1}} \geq c_3 \,  |V_k \cap
B_{1 \over 2}|^{n-1 \over n} \geq c_3  \, 2^{1-n \over n} \,
|V_k|^{n-1 \over n} \, ,
$$
where $c_3=c_3({\mathbb R}^n, \gamma/2 )$.
\par
{\it Case 2b.} We suppose that for all $k$
\begin{equation}\label{27}
|V_k  \setminus B_{1 \over 2} |  \geq {1 \over 2} \, |V_k| \, .
\end{equation}
Denote
$$
\omega (x) := {x \over |x|^2} \,
$$
and
\begin{equation}\label{16}
V_k^0:=\omega \left(V_k \setminus B_{1 \over 2} \right) \, ,
\qquad A_k^0 := \omega \left(A_k \setminus B_{1 \over 4} \right)
\, ,
\end{equation}
$\widehat V_k=V_k^0 \cup V_k$ and  $\widehat A_k=A_k^0 \cup A_k$.
Clearly, $\widehat V_k \subset \widehat A_k \subset B_4$ and
\begin{equation}\label{19}
(B_4 \setminus B_1) \setminus A_k^0= \omega \left((B_1 \setminus
B_{1 \over 4}) \setminus A_k \right) \, .
\end{equation}
The elementary calculations show that
\begin{equation}\label{22}
|\nabla \omega (x)| ={1 \over |x|^{2n}}
\end{equation}
and
\begin{align} \label{23}
|\omega (x)-\omega(y)| &\geq |x-y| \qquad \textrm{if} \quad |x|,
|y| \leq 1
\\
\label{32}
|x-\omega(y)| &\geq |x-y| \qquad \textrm{if} \quad |x|,
|y| \leq 1 \, .
\end{align}
Making the change of  variables $x:=\omega (x)$, $y:=\omega (y)$
and applying~\reff{16}, \reff{19}, \reff{23}  we obtain
\begin{align}\nonumber
\int\limits_{(B_4 \setminus B_1) \setminus A_k^0} \, \,  \, \,
\int\limits_{V_k^0} {dx dy \over |x-y|^{n+1}}  &=
\int\limits_{(B_1 \setminus B_{1 \over 4}) \setminus A_k}
 \, \,  \, \,
\int\limits_{V_k \setminus B_{1 \over 2}} {|\nabla \omega (x) | \,
|\nabla \omega (y) | dx dy \over |\omega (x)-\omega(y)|^{n+1}}
\\
\nonumber &\leq 8^{2n} \int\limits_{(B_1 \setminus B_{1 \over 4})
\setminus A_k}
 \, \,  \, \,
\int\limits_{V_k \setminus B_{1 \over 2}} { dx dy \over
|x-y|^{n+1}}
\\
\label{36} &\leq 8^{2n} \int\limits_{B_1 \setminus A_k}
 \, \,  \, \,
\int\limits_{V_k } { dx dy \over |x-y|^{n+1}} \, ,
\end{align}
where we have used, by~\reff{22},
$$
|\nabla \omega (x)| \leq 4^{2n} \quad \textrm{for} \quad x \in B_1
\setminus B_{1 \over 4} \, ,  \qquad |\nabla \omega (y)| \leq
2^{2n} \quad \textrm{for} \quad y \in B_1 \setminus B_{1 \over 2}
\, .
$$
Similarly, making the change of  variables and
using~\reff{16}-\reff{22}, \reff{32} we have
\begin{equation}\label{34}
\int\limits_{(B_4 \setminus B_1) \setminus A_k^0}  \, \,  \, \,
\int\limits_{V_k} {dx dy \over |x-y|^{n+1}}  \leq 4^{2n}
\int\limits_{B_1 \setminus A_k}  \, \,  \, \, \int\limits_{V_k } {
dx dy \over |x-y|^{n+1}}
\end{equation}
and
\begin{equation}\label{35}
\int\limits_{B_1 \setminus A_k}  \, \,  \, \, \int\limits_{V_k^0}
{dx dy \over |x-y|^{n+1}} \leq 2^{2n} \int\limits_{B_1 \setminus
A_k}  \, \,  \, \, \int\limits_{V_k } { dx dy \over |x-y|^{n+1}}
\, .
\end{equation}
Combining~\reff{36}-\reff{35} we arrive at
\begin{equation}\label{31}
\int\limits_{B_4 \setminus \widehat A_k}  \, \,  \, \,
\int\limits_{\widehat V_k} {dx dy \over |x-y|^{n+1}} \leq
(1+2^{2n}+4^{2n}+8^{2n}) \int\limits_{B_1 \setminus A_k}  \, \, \,
\,  \int\limits_{V_k } { dx dy \over |x-y|^{n+1}} \, .
\end{equation}
Since $|x-y| \leq 6$ for $x \in B_4 \setminus B_3$ and $y \in
B_2$, it follows that
$$
\int\limits_{B_4 \setminus \widehat A_k} {dx  \over |x-y|^{n+1}}
\geq \int\limits_{(B_4 \setminus B_3) \setminus  A_k^0} {dx \over
|x-y|^{n+1}}  \geq {1 \over 6^{n+1}} \, |(B_4 \setminus B_3)
\setminus  A_k^0|
$$
for all $y \in B_2$. In view of~\reff{24} and~\reff{22},
$$
|(B_4 \setminus B_3) \setminus  A_k^0| =|\omega \left((B_{1 \over
3} \setminus B_{1 \over 4}) \setminus A_k \right)| \geq |(B_{1
\over 3} \setminus B_{1 \over 4}) \setminus A_k | \geq {1 \over 2}
|B_{1 \over 3} \setminus B_{1 \over 4}| \, .
$$
Thus
$$
\int\limits_{B_4 \setminus \widehat A_k} {dx  \over |x-y|^{n+1}}
\geq c_7 >0 \, ,
$$
where $c_7=6^{-n-2}|B_{1 \over 3} \setminus B_{1 \over 4}|$.
Moreover, for any $y \in B_2$
$$
\int\limits_{{\mathbb R}^n \setminus B_4} {dx \over |x-y|^{n+1}}
\leq \int\limits_{{\mathbb R}^n \setminus B_4} {dx \over
(|x|-2)^{n+1}}=:c_8 \, .
$$
Thus for all $y \in B_2$  one has
\begin{equation}\label{30}
\int\limits_{B_4 \setminus \widehat A_k} {dx  \over |x-y|^{n+1}}
\geq c_9 \int\limits_{{\mathbb R}^n \setminus \widehat A_k} {dx
\over |x-y|^{n+1}} \, ,
\end{equation}
where $c_9=c_7 /(c_8+c_7)$. Combining~\reff{31}, \reff{30} and
using $\widehat V_k \subset B_2$ we arrive at
\begin{equation}\label{39}
\int\limits_{B_1 \setminus A_k}\, \, \, \,  \int\limits_{V_k } {
dx dy \over |x-y|^{n+1}} \geq  c_{10} \int\limits_{{\mathbb R}^n
\setminus \widehat A_k} \, \, \, \,  \int\limits_{\widehat V_k }
{dx dy \over |x-y|^{n+1}} \, ,
\end{equation}
where $c_{10}=c_9 /(1+2^{2n}+4^{2n}+8^{2n})$. Observe that
$$
|V_k^0| \geq |V_k \setminus B_{1\over 2}| \geq {1 \over 2} \,
|V_k| \, ,  \qquad |A_k^0| \leq 4^{2n} |A_k \setminus B_{1 \over
4}| \leq 4^{2n} |A_k| \, .
$$
Consequently,
$$
|V_k^0| \geq {1 \over 2} \, |V_k| \geq {\gamma \over 2} \, |A_k|
\geq \gamma 4^{-2n-1} |A_k^0|
$$
and so
$$
|\widehat V_k| \geq \gamma 4^{-2n} |\widehat A_k| \,  .
$$
Using this we apply Lemma~\ref{l1} with $\Omega={\mathbb R}^n$,
$A=\widehat A_k$, $V=\widehat V_k $ to get
$$
\int\limits_{{\mathbb R}^n \setminus \widehat A_k}
\int\limits_{\widehat V_k } {dx dy \over |x-y|^{n+1}} \geq
c_3({\mathbb R}^n, \gamma 4^{-2n}) \, |\widehat V_k |^{n-1 \over
n} \geq c_3({\mathbb R}^n, \gamma 4^{-2n}) |V_k|^{n-1 \over n} \,
.
$$
This and~\reff{39} prove Lemma~\ref{l1} for Case 2b.
\par
{\it Case 3.} Let $\Omega=\phi (B_1)$, where $\phi$ satisfies the
conditions of Theorem~\ref{t1}. Making the change of  variables
$x:=\phi(x)$, $y:=\phi (y)$ we get
$$
\int\limits_{\Omega \setminus A} \int\limits_{V} {dx dy \over
|x-y|^{n+1}} = \int\limits_{B_1 \setminus \widetilde A}
\int\limits_{\widetilde V} {|\nabla \phi (x)| \, |\nabla \phi
(y)|\, dx dy \over |\phi(x)-\phi(y)|^{n+1}} \, ,
$$
where $\widetilde A = \phi^{-1} (A)$, $\widetilde V = \phi^{-1}
(V)$. Because of~\reff{29} it follows that
$$
|\nabla \phi(x)| \geq c_1^{-1} \, .
$$
Hence
$$
\int\limits_{\Omega \setminus A} \int\limits_{V} {dx dy \over
|x-y|^{n+1}} \geq c_1^{-(n+3)} \int\limits_{B_1 \setminus
\widetilde A} \int\limits_{\widetilde V} { dx dy \over
|\phi(x)-\phi(y)|^{n+1}} \, .
$$
Since $|\widetilde V| \geq c_1^{-1} |V| \geq  c_1^{-1} \gamma |A|
\geq c_1^{-2} \gamma |\widetilde A|$, an application of
Lemma~\ref{l1} with $(\Omega, A, V):=(B_1,\widetilde A,\widetilde
V) $ gives
$$
\int\limits_{\Omega \setminus A} \int\limits_{V} {dx dy \over
|x-y|^{n+1}} \geq c_1^{-(n+3)} c_3  \, |\widetilde V| \geq
c_1^{-(n+4)} c_3 \, |V| \, ,
$$
where $c_3=c_3 (B_1, c_1^{-2} \gamma )$.
\end{proof}
\begin{Lemma}\label{l2}
Let $\Omega \subset {\mathbb R}^n$, $n \geq 2$ and suppose that
$f:\Omega \rightarrow {\mathbb C}$ be a continuous function, which
satisfies conditions of Theorem~\ref{t1}. Let
\begin{equation}\label{4}
I_x:=\{y \in \Omega : |f(x)| > 2 |f(y)|\} \, , \qquad J_x:=\{y \in
\Omega : |f(y)| > 2 |f(x)|\}
\end{equation}
and
\begin{equation}\label{9}
\psi (x) = \int\limits_{I_x \cup J_x} \, {dy \over |x-y|^{n+1}} \,
.
\end{equation}
 Then
\begin{equation}\label{10}
\int\limits_{\Omega} |f (x)|^2  \psi(x) \, dx +
\int\limits_{\Omega}|f (x)|^2 \, dx \geq c_{11} \left(
\int\limits_{\Omega} |f (x)|^{2n \over n-1} \, dx
\right)^{n-1\over n} \,
\end{equation}
for some  $c_{11}=c_{11} (\Omega) >0$.
\end{Lemma}
\begin{proof}
Without loss of generality we may assume that $|f(x)| \leq 1$ for
all $x \in \Omega$.
\par
For any $m \in {\mathbb N}$ we put
$$
D_m := \{x \in \Omega: f(x) \in [ 2^{-m},  2^{-m+1}]\} \,
$$
and $D_\infty=f^{-1}  (0) $. Clearly, $\Omega=D_\infty \cup
\bigcup\limits_{m=1}^\infty D_m$. One has
\begin{equation}\label{20}
\int\limits_{\Omega} |f(x)|^{2n \over n-1} \, dx \leq
\sum\limits_{m=1}^\infty {|D_m| \over N^{m-1}} \, ,
\end{equation}
where $N=2^{2n \over n-1}$. Denote by $E \subset {\mathbb N}$ the
set of indexes $m$ such that
\begin{equation}\label{13}
|D_m| \geq {|D_{m-1}|+ |D_{m+1}| \over 64} \, .
\end{equation}
Since
\begin{align*}
{|D_{m-1}|+ |D_{m+1}| \over  64 N^{m-1}} &\leq {\max\{N, N^{-1}\}
\over 64} \left({|D_{m-1}| \over  N^{m-2}}+ {|D_{m+1}| \over
N^{m}} \right) \\ &\leq {1 \over 4} \left({|D_{m-1}| \over
N^{m-2}}+ {|D_{m+1}| \over   N^{m}} \right) \, ,
\end{align*}
it follows that
\begin{align*} \sum\limits_{m=1}^\infty {|D_m| \over
N^{m-1}} &= \sum\limits_{m \in E}{|D_m| \over
N^{m-1}}+\sum\limits_{m \in {\mathbb N} \setminus E}{|D_m| \over
N^{m-1}} \\ &\leq \sum\limits_{m \in E}{|D_m| \over N^{m-1}}+
\sum\limits_{m \in {\mathbb N} \setminus E}{|D_{m-1}|+ |D_{m+1}|
\over  64 N^{m-1}}
\\
&\leq \sum\limits_{m \in E}{|D_m| \over N^{m-1}}+ {1 \over 2} \,
\sum\limits_{m =1}^\infty {|D_{m}| \over  N^{m-1}} \, ,
\end{align*}
where we have used that $64 |D_m| < |D_{m-1}|+ |D_{m+1}|$ for $m
\in {\mathbb N} \setminus E$. Hence
\begin{equation}\label{21}
\sum\limits_{m=1}^\infty {|D_m| \over N^{m-1}} \leq 2
\sum\limits_{m \in E}{|D_m| \over N^{m-1}} \, .
\end{equation}
Consequently, from~\reff{20} and~\reff{21} we deduce that
\begin{equation}\label{8}
\left(\int\limits_{\Omega} |f(x)|^{2n \over n-1} \, dx
\right)^{n-1 \over n}  \leq 2^{n-1 \over n} \sum\limits_{m \in E}
{ |D_m|^{n-1 \over n} \over 4^{m-1}} \leq 8\,  \sum\limits_{m \in
E} { |D_m|^{n-1 \over n} \over 4^{m}} \, ,
\end{equation}
where we have used that $N^{n-1 \over n}=4$ and the elementary
inequality
\begin{equation}\label{7}
\left(\sum\limits_{k=1}^\infty a_k\right)^{n-1 \over n} \leq
\sum\limits_{k=1}^\infty a_k^{n-1 \over n}
\end{equation}
for all $a_k \geq 0$.
\par
We split the remaining part of the proof into two cases.
\par {\it Case 1.} Suppose that there exists $p \geq 2$ such that
\begin{equation}\label{14}
|D_{p-1}|+|D_{p}|+|D_{p+1}| > \sum\limits_{m=1 \atop m \not = p-1,
p, p+1} |D_m| \, .
\end{equation}
From definition~\reff{4} we see that $\bigcup\limits_{k=1}^{m-2}
D_k \subset J_x$ and $D_\infty \cup \bigcup\limits_{k=m+2}^\infty
D_k \subset I_x$. Using~\reff{9} we obtain
\begin{align}\nonumber
\int\limits_{\Omega} |f(x)|^2 \psi (x) \, dx &\geq
\sum\limits_{m=1 }^\infty \int\limits_{D_m} \int\limits_{I_x \cup
J_x} |f(x)|^2 { dx dy \over |x-y|^{n+1}} \\
\label{6} &\geq \sum\limits_{m=1 \atop m \not = p-2,\ldots,
p+2}^\infty {1 \over 4^{m}} \int\limits_{D_m} \int\limits_{L_m} {
dx dy \over |x-y|^{n+1}} \, ,
\end{align}
where
\begin{equation}\label{1}
L_m:=D_\infty \cup \bigcup\limits_{k=1 \atop k \not = m-1, m ,
m+1}^{\infty} D_k \, , \qquad m \in {\mathbb N}\, .
\end{equation}
From~\reff{14} we find
\begin{equation}\label{11}
|L_m| \geq |D_{p-1}|+|D_{p}|+|D_{p+1}| \geq {|\Omega| \over 2}
\end{equation}
for $m \not = p-2, p-1, p, p+1, p+2$. Hence ~\reff{13}
and~\reff{11} imply~\reff{17} and~\reff{5} respectively with
$A=D_{m-1} \cup D_m \cup D_{m+1}$,  $V=D_m$ and $\gamma=1/65$. An
application of Lemma~\ref{l1} gives
$$
\int\limits_{D_m} \int\limits_{L_m} { dx dy \over |x-y|^{n+1}}
\geq c_3  \, {|D_m|^{n-1 \over n} \over 4^m} \, ,
$$
where $c_3=c_3(\Omega, 1/65)$. Therefore
\begin{equation}\label{3}
\int\limits_{\Omega} |f(x)|^2 \psi (x) \, dx \geq c_3 \,
\sum\limits_{m=1 \atop m \not = p-2,\ldots, p+2}^\infty
{|D_m|^{n-1 \over n} \over 4^m} \geq c_3 \, \sum\limits_{m \in E
\atop m \not = p-2, \ldots , p+2}  {|D_m|^{n-1 \over n} \over 4^m}
\, .
\end{equation}
\par
Using the second inequality in~\reff{11} we have
\begin{align}\nonumber
\int\limits_{\Omega} |f(x)|^2 \, dx \geq \sum\limits_{m=p-2}^{p+2}
{|D_m| \over 4^{m}}  \geq {\sum\limits_{m=p-2}^{p+2}  |D_{m}|
\over 4^{p+2}}  &\geq \left( {|\Omega|\over 2}\right)^{1 \over n}
{\left(\sum\limits_{m=p-2}^{p+2}  |D_{m}| \right)^{n-1 \over n}
\over 4^{p+2}}
\\
 \nonumber
&\geq {1 \over 5} \, \left( {|\Omega|\over 2}\right)^{1 \over n}
{\left(\sum\limits_{m=p-2}^{p+2}  |D_{m}|^{n-1 \over n}\right)
\over 4^{p+2}}
\\
\label{12} &\geq c_{12}\sum\limits_{m=p-2}^{p+2}   {|D_{m}|^{n-1
\over n}
 \over 4^{m}} \, ,
\end{align}
where
$$
c_{12} =\min\left\{c_3 , {1 \over 4^4} \, {1 \over 5 } \, \left(
{|\Omega|\over 2}\right)^{1 \over n} \right\} \, .
$$
Piecing together~\reff{3} and~\reff{12} we have
$$
\int\limits_{\Omega} |f(x)|^2 \psi (x) \, dx +\int\limits_{\Omega}
|f(x)|^2  \, dx \geq c_{12} \, \sum\limits_{m \in E} {|D_m|^{n-1
\over n} \over 4^m} \,.
$$
Combining this and~\reff{8} we arrive at~\reff{10} with
$c_{11}=c_{12}/8$.
\par
{\it Case 2.} Let us assume that
$$
|D_{m-1}|+|D_{m}|+|D_{m+1}| \leq  \sum\limits_{k=1 \atop k \not =
m-1, m, m+1} |D_k| \,
$$
for all $m \geq 2$. Then for all $m$ we have
$$
|L_m| \geq {|\Omega| \over 2} \, ,
$$
where $L_m$ is defined by~\reff{1}. As in Case 1 we apply
Lemma~\ref{l1} to get
\begin{equation}\label{2}
\int\limits_{\Omega} |f(x)|^2 \psi (x) \, dx  \geq c_3 \,
\sum\limits_{m \in E }  {|D_m|^{n-1 \over n} \over 4^m} \, .
\end{equation}
Comparing~\reff{8} and~\reff{2} finishes the proof.
\end{proof}
\begin{proof}[Proof of Theorem~\ref{t1}.]
As  was shown in \cite{BT} \mbox{(consequence of Theorem 2 \,
p.15)} for any domain $\Omega$ satisfying conditions of
Theorem~\ref{t1} and for any continuous function $f: \Omega
\rightarrow {\mathbb C}$ we have
\begin{equation}\label{37}
\int\limits_{\Omega \times \Omega} {|f(x)-f(y)|^2 \over
|x-y|^{n+1}} \, dxdy \geq c_{13}  \int\limits_{\Omega} |f(x)|^2 \,
dx \,
\end{equation}
for some constant $c_{13}=c_{13} (\Omega)>0$. On the other hand
$$
\int\limits_{\Omega \times \Omega} {|f(x)-f(y)|^2 \over
|x-y|^{n+1}} \, dxdy \geq \int\limits_{\Omega} \left( \, \,
\int\limits_{I_x \cup J_x} {|f(x)-f(y)|^2 \over |x-y|^{n+1}} \, dy
\right) \, dx \, ,
$$
where $I_x$ and $J_x$ are given by~\reff{4}. Since
$$
|f(x)-f(y)|\geq {|f(x)| \over 2}
$$
for all $y \in I_x \cup J_x$, it follows that
\begin{align}\nonumber
\int\limits_{\Omega \times \Omega} {|f(x)-f(y)|^2 \over
|x-y|^{n+1}} \, dxdy &\geq {1 \over 4} \, \int\limits_{\Omega}
|f(x)|^2 \left( \, \, \int\limits_{I_x \cup J_x} {dy \over
|x-y|^{n+1}} \right) \, dx
\\
\label{38} &={1 \over 4} \, \int\limits_{\Omega} |f(x)|^2 \psi (x)
\, dx \, .
\end{align}
Summing~\reff{37}, \reff{38} and applying Lemma~\ref{l2} completes
the proof.
\end{proof}

{\bf Acknowledgements}. We are grateful for financial support from
\mbox{EPSRC} Grant RCMT090. We also thank Professor W.D. Evans for
valuable discussions.



\end{document}